\documentclass[english,keywords,showpacs,amsmath,amssymb,twocolumn]{revtex4-1}
\usepackage[T1]{fontenc}
\usepackage[latin1]{inputenc}
\usepackage{babel}
\usepackage{array}
\usepackage{graphics}
\makeatother
\begin{document}
\title{Reexamining Larmor precession in a spin-rotator: Testable correction and its ramifications}

\author{Dipankar Home}
\altaffiliation{dhome@bosemain.boseinst.ac.in}
\affiliation{CAPSS, Departmrnt of Physics, Bose Institute, Salt Lake, Kolkata-700091, India}

\author{Alok Kumar Pan}
\altaffiliation{akp@math.cm.is.nagoya-u.ac.jp}
\affiliation{Graduate School of Information Science, Nagoya University, Chikusa-ku, Nagoya 464-8601, Japan}

\author{Arka Banerjee}
\altaffiliation{arka@theory.tifr.res.in}
\affiliation{Tata Institute of Fundamental Research, Homi Bhabha Road, Mumbai-400005, India}

\begin{abstract}
For a spin-polarized plane wave passing through a spin-rotator containing  uniform magnetic field, we provide a detailed analysis for solving the appropriate Schr\"{o}dinger equation.  A modified expression for spin precession is obtained which reduces to the standard Larmor precession relation when kinetic energy is very large compared to the spin-magnetic field interaction. We show that there are experimentally verifiable regimes of departure from the standard Larmor precession formula. The treatment is then extended to the case of a spin-polarized wave packet passing through a uniform magnetic field. The results based on the standard expression for Larmor precession and that obtained from the modified formula are compared in various regimes of the  experimental parameters.
\end{abstract}
\maketitle

\section{Introduction}
If a spin-1/2 particle passes through a region of uniform magnetic field, it is well known that the time evolution of the spin of the particle undergoes what is commonly known as Larmor precession. For example, if the particle with an initial spin orientation along the $+\hat x$ axis passes through a magnetic field oriented along the +$\hat z$ axis, the spin precesses in the x-y plane with a frequency determined by the strength of the magnetic field and the magnetic moment of the particle. This frequency is known as the Larmor frequency - a commonly discussed topic that has wide applications\cite{mezei1,zei,mezei2,alefeld,heb,otake,hino,rek,mar}, particularly in the analysis of experiments involving neutron, electron and atomic interferometry and in calculating the tunneling time \cite{baz67a,ryb,buttiker82,buttiker83,falck88} through a potential barrier. 

The usual treatments on the standard Larmor precession essentially consider a situation where the particle is stationary, trapped within a region containing uniform magnetic field \cite{landau,sakurai,merz,cohen,grif,greiner,bohmbook}, thereby ignoring the spatial part of the wave function and the time evolution of the wave function is considered only in terms of the potential energy arising out of the spin-magnetic field interaction. The argument for ignoring the kinetic energy term in the Hamiltonian seems to take this term to be much smaller in magnitude as compared to the spin-magnetic field potential energy term. In our treatment, the incident spin-1/2 particles are considered to be passing through a spatial region within which a constant magnetic field is confined so that in treating the time evolution, both the kinetic energy term and the spin-magnetic field potential energy term are taken into account. Interestingly, the Larmor precession relation is recovered when the spin-magnetic field interaction energy is  much smaller compared to the kinetic energy term. In this case, the path and spin degrees of freedom can then be treated independently.

We would also like to note that usual treatments of calculating tunneling time through a barrier based on Larmor clock \cite{baz67a,ryb,buttiker82,buttiker83,falck88}  pertain to a spatial region within which a constant magnetic field and an external potential $V_{0}$  are both confined so that both spin-up and spin-down particles effectively see potential barriers of different heights. On the other hand, in our treatment, we do not consider any external potential so that while a spin-up particle sees a potential barrier, a spin-down particle sees a well, and eventually a path-spin entangled state is generated.  To the best of our knowledge, it is only in the Appendix of the treatment given by Buttiker \cite{buttiker83} that the question of when the Larmor precession relation is valid in the absence of any external potential barrier is briefly discussed. Here we give a detailed analysis of this issue, providing a clear delineation of the regime of deviation from the standard Larmor precession relation.  

We begin by considering a wave function whose space part is a plane wave and is spin polarized in the $+x$ direction. By examining closely the time evolution of the entire wave function, that is, both the spin and the space parts, caused by the interaction of the spin of the particle with the magnetic field, we find an interesting feature. Due to the specifics of the spin-magnetic field interaction, it is possible to derive the time evolution of the entire wave function by solving the time-independent Schrodinger equation for only the spatial part. Then, from the time-evolved entire wave function, one can find the change in the spin part of the wave function; this is shown in Section II. Subsequently, in Section III, the limit in which the result of our treatment matches the result of the standard Larmor precession as well as the limit in which there is an appreciable departure from Larmor precession are discussed. This treatment reveals that it is, in fact, the limit where the kinetic energy is much higher than the potential energy due to the spin-field interaction that the standard expression for Larmor precession holds true. In Section III numerical estimates of departure from the standard Larmor precession are presented. In Section IV we generalize the treatment given above to the case of an incident wave packet.

\section{Spin-rotator containing a uniform magnetic field}
We consider particles passing through a spin-rotator containing a constant magnetic field directed along  the $+z$-axis  in a region between $x=0$ and $x=a$. The total incident wave function of a particle is represented by $\Psi_{i}=\psi_{0}\otimes\chi$ , where  $\psi_0=A e^{i k x}$ is the spatial part taken to be a plane wave with wave number $k$, and  $\chi=(\frac{1}{\sqrt{2}}\left(\left|\uparrow \right\rangle_{z}+ \left|\downarrow\right\rangle_{z}\right)$ is the spin state polarized in the $+x$ direction with $|\uparrow\rangle_{z}$ and $|\downarrow\rangle_{z}$ are the eigenstates of $\widehat{\sigma}_{z}$. Hence, our setup is different from the case where the particle is trapped within the region containing uniform magnetic field.

The interaction Hamiltonian is $H_{int}=-\mu_n\vec {\sigma}.\textbf{B}$ where $\mu_n$ is the magnetic moment of the neutron, $\textbf{B}= B\widehat{z}$ is the homogeneous magnetic field and $\vec{\sigma}$ is the Pauli spin vector. Since $\mu_n$ is known to be a negative quantity, it is convenient to define for further calculations a quantity $\mu=-\mu_n$.  Here note that the magnetic field has an implicit position dependence as it is confined between $x=0$ and $x=a$. In this case, the two-component Pauli equation can be written as the following two coupled
equations for the time evolution of the spatial parts $\psi^+$ and $\psi^-$, corresponding to the spin $|\uparrow\rangle_{z}$ and $|\downarrow\rangle_{z}$ components respectively  

\begin{equation}
\label{pauli1}
i\hbar\frac{\partial\psi^{+}}{\partial t}=-\frac{\hbar^{2}}{2m}\nabla^{2}\psi^{+}+\mu  B\widehat{z}\psi^{+}
\end{equation}
\begin{equation}
\label{pauli2}
i\hbar\frac{\partial\psi^{-}}{\partial t}=-\frac{\hbar^{2}}{2m}\nabla^{2}\psi^{-} - \mu B\widehat{z}\psi^{-}
\end{equation}
Eqs.(\ref{pauli1}) and (\ref{pauli2}) imply that while a neutron having spin-up interacts with the spin rotator containing constant magnetic field, its associated spatial wave function $(\psi^{+})$ evolves under a \emph{potential barrier} that has been generated due to the spin-magnetic field interaction; on the other hand, the associated spatial wave function $(\psi^{-})$ for a spin-down neutron evolves under a \emph{potential well}. 

\begin{figure}[t]
{\rotatebox{0}{\resizebox{8.0cm}{5.5cm}{\includegraphics{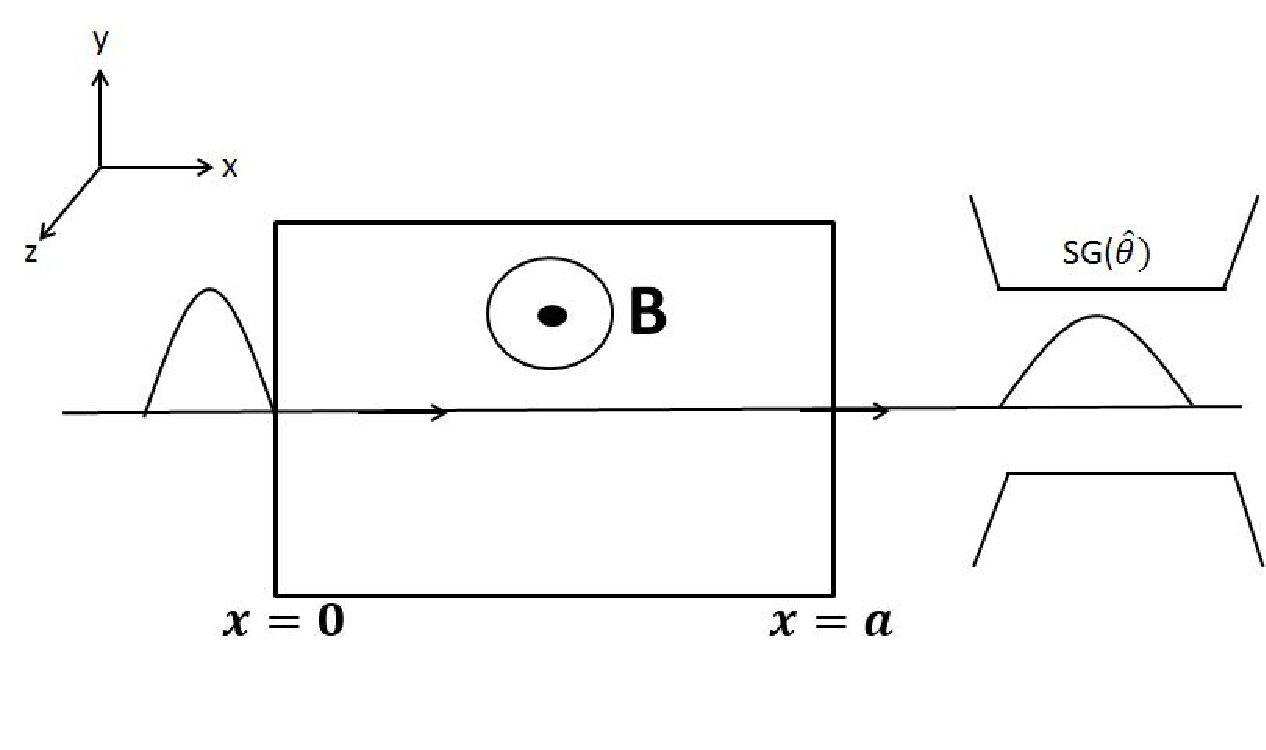}}}}
\caption{\label{fig.1} {\footnotesize Spin-1/2 particles with initial spin orientations polarised along the \( +\widehat{\bf x} \)
- axis pass through a spin-rotator (SR) containing a constant magnetic field \( {\bf B} \) directed along the \( +\widehat{\bf z} \)
- axis. The particles emerging from the SR have a distribution of
their spins oriented along different directions. Calculation of this
distribution function is experimentally tested by measuring the spin
observable along a direction \( \widehat{n}\left( \theta \right) \) by suitably orienting the direction
\( \widehat{n}\left( \theta \right)  \) of the inhomogeneous magnetic
field in the Stern-Gerlach \( \left( SG-\widehat{n}\right)  \) device.}}
\end{figure}
Then the time evolved total wave function  at $t=\tau$ after the interaction of 
spins with the uniform magnetic field is given by
\begin{eqnarray}
\label{entstate}
\nonumber
\Psi\left(\textbf{x},\tau\right) &=& \exp({-\frac{iH\tau}{\hbar}})\Psi(\textbf{x},0)\\
&=&\frac{1}{\sqrt{2}}\left[\psi^{+}(\textbf{x},\tau)\otimes\left|\uparrow\right\rangle_{z}+\psi^{-}(\textbf{x}, \tau)\otimes\left|\downarrow\right\rangle_{z}\right]
\label{timeevolved}
\end{eqnarray}
where $\psi^{+}\left({\bf x},\tau\right)$ and 
$\psi^{-}\left({\bf x},\tau\right)$
are the two components of the spinor 
$\psi=\left(\begin{array}{c}\begin{array}{c} 
\psi^{+}\\ \psi^{-}\end{array}\end{array}\right)$ which satisfies the 
Pauli equation. The homogeneous magnetic field is written as 
${\bf B}=B \widehat{z}$.

Note that the entanglement between the spin and the spatial parts of the wave function embodied in Eq.(\ref{entstate})results from the application of Pauli equation in this problem using the spin-magnetic field interaction term which has an implicit spatial dependence of the confinement of the uniform magnetic field within the bounded region of the spin rotator. It is this feature taken into account in the treatment given in this paper that leads to a nontrivial modification of the standard formula for Larmor precession.

Before proceeding to focus on the exact solutions of Eqs. (\ref{pauli1}) and (\ref{pauli2}), we revisit in the next section the usual treatment of this problem  used to derive the Larmor precession relation for the rotated spin state after the spin-magnetic field interaction within the spin-rotator.

\subsection{The usual treatment of Larmor precession}
We may note here again that the usual treatments pertain essentially to a stationary particle within a spin-rotator. The behavior of the wave function after it starts interacting with the spin rotator is usually described by taking into account \textit {only} the spin part of the wave function, while the space part is completely left out of the analysis\cite{landau,sakurai,merz,cohen,grif,greiner,bohmbook}. Neglecting the kinetic energy, the Hamiltonian of the system inside the spin rotator is taken to be $H= \mu\vec{\sigma}.\bf{B}$.  The spin up and spin down parts of the wave function evolve according to the Schroedinger equation in the following way
\begin{equation}
\label{so+}
i\hbar\frac{\partial \psi^{+}}{\partial t}=  \mu B \psi^{+}
\end{equation}
\vskip -0.5cm
\begin{equation}
\label{so-}
i\hbar\frac{\partial \psi^{-}}{\partial t}= -\mu B \psi^{-}
\end{equation}
The  solutions of the above two first order differential equations are $\psi^{\pm}=\psi_{0}e^{\mp i \omega \tau/2}$ where $\omega= 2\mu B/\hbar$, and $\tau$ is the time over which the spin-magnetic field interaction takes place.  Putting these solutions in  Eq.(\ref{entstate}), we finally obtain
\begin{eqnarray}
\label{psitauusual}
\Psi(x,\tau)=\psi_{0} \frac{1}{\sqrt{2}}\left(e^{- i \omega \tau}\left|\uparrow\right\rangle+ e^{i \omega \tau}\left|\downarrow\right\rangle\right)
\end{eqnarray}
Eq.(\ref{psitauusual}) can be written as
\begin{eqnarray}
\Psi(x,\phi)=\psi_{0}\frac{e^{-i\phi/2}}{\sqrt{2}}\left(\left|\uparrow\right\rangle_{z}+ e^{i\phi}\left|\downarrow\right\rangle_{z}\right)\equiv e^{- i\phi/2}\psi_{0} \chi(\phi)
\end{eqnarray} 
where
\begin{equation}
\phi=\omega \tau
\end{equation}
is the well-known Larmor precession relation and $\omega $ is the Larmor frequency and, $e^{- i\phi/2}$ is the global phase. The spin state after the interaction is given by $\chi(\phi)=\frac{1}{\sqrt{2}}\left(\left|\uparrow\right\rangle+ e^{i\phi}\left|\downarrow\right\rangle\right)$. 

In the above treatment $\tau$ is taken to be the transit time through the spin rotator, given by $\tau=\frac{a}{v}$ where $a$ is the spatial extension of spin-rotator containing the uniform magnetic field, and $v=\frac{\hbar k}{m}$ is the initial velocity of particle.  

The detailed analysis of this problem for particles passing through the spin rotator should include the evolutions of both the spin and space parts of the total incident wave function. In the following section we provide such a treatment. 

\subsection{The detailed analysis for particles passing through a spin-rotator}
We first consider the spatial part of the total incident wave function to be a plane wave corresponding to a single wave vector.  Later, we shall consider the spatial part as a wave packet with superposition of plane waves. 

Here we begin by recalling that Eqs. (\ref{pauli1}) and (\ref{pauli2}) imply that in this problem, we effectively have a situation where the spin up part of the wave function faces a \emph{potential barrier} while the spin down part of the wave function faces a \emph{potential well}. This, in turn, entails that the information about the spin part of the wave function enters the space part of the wave function through this potential, and thus solving the Schroedinger equation for only the space part of the wave function suffices to get complete information about the time evolution of the combined spin-space wave function. Therefore, instead of equations (\ref{so+}) and (\ref{so-}), one needs to solve Eqs.(\ref{pauli1}) and (\ref{pauli2}) explicitly. 

The solutions to these equations, given our incident state, consist of a reflected part traveling in the $-x$-direction and a transmitted part of the wave function, traveling in the $+x$-direction. We should note here that the reflected part of the wave function exists \textit {only} to the left of the spin rotator, and the transmitted part exists \textit {only} to the right of the spin rotator. Our ultimate objective is to calculate  the observable rotation of the spin part of the wave function caused by evolution of the state due to the spin-magnetic field interaction within the spin rotator.

For this, we need to look only at that part of the wave function which pertains to those neutrons which have actually passed through the spin rotator, or in other words we focus only on the transmitted part of the wave function. Therefore, using the solutions of equations (\ref{pauli1}) and (\ref{pauli2}), we will finally end up with the following state
\begin{equation}
\label{fitra}
\left|\Psi_{f}\right\rangle=\frac{N}{\sqrt{2}} ( \psi^{+}_T\left|\uparrow\right\rangle + \psi^{-}_{T}\left|\downarrow\right\rangle)
\end{equation}
where $N$ is the normalized constant can be written as $N=\int_{v}(|\psi^{+}_T|^{2} + |\psi^{-}_T|^{2}) dv$.\\

Note that, Eq.(\ref{fitra}) is an entangled state between the spin and spatial degrees of freedom of the transmitted part. The expressions for the reflected and transmitted parts of a wave function at a potential well or barrier are well known. However, from an empirical perspective, we note that if in the above setup, even if we use low energy or ultra-cold neutrons, having kinetic energy of the order of $5 \times 10^{-7} eV$, for the potential energy term ($|\mu B|$) to exceed the kinetic energy term ($E$), we will need a field of the order of $10 T$. Magnetic fields of such high intensity are difficult to produce in laboratory conditions, and therefore for all practical purposes, we should consider the situation where $E>\mu B$. 

Now, since $\psi^{-}$ evolves under a potential well confined between $x=0$ and $x=a$ the reflected and transmitted parts are respectively given by
\begin{equation}
\label{psi-r}
\psi^{-}_{R}= A e^{- i k x}\frac{(k^2- k_1^2)(1-e^{2 i k_1 a})}{(k + k_1)^2- (k- k_1)^2 e^{2 i k_1 a} }; \quad x<0
\end{equation}

\begin{equation}
\label{psi-t}
\psi^{-}_{T}= A e^{i k x}\frac{4 k k_1 e^{- i k a} e^{i k_1 a}}{(k + k_1)^2- (k- k_1)^2 e^{2 i k_1 a} }; \quad x>a
\end{equation}
where $k=\frac{\sqrt{2 m E}}{\hbar}$ , $k_1= \frac{\sqrt{2 m (E+\mu B)}}{\hbar}$ and $a$ is the width of the spin rotator arrangement which contains the uniform magnetic field. 
The wave function $\psi^{+}$ evolves under a potential well, the expressions for the transmitted and the reflected part by replacing all the $k_1$'s in  Eqs.(\ref{psi-r}) and (\ref{psi-t}) by $k_2$ where $k_2=\frac{\sqrt{2 m (E-\mu B)}}{\hbar}$.
\begin{equation}
\psi^{+}_{R}= A e^ {-i k x}\frac{(k^2- k_2^2)(1-e^{2 i k_2 a})}{(k + k_2)^2- (k- k_2)^2 e^{2 i k_2 a} }; \quad x<0
\end{equation}

\begin{equation}
\psi^{+}_{T}= A e^{i k x}\frac{4 k k_2 e^{- i k a} e^{i k_2 a}}{(k + k_2)^2- (k- k_2)^2 e^{2 i k_2 a} }; \quad x>a
\end{equation}
Then, in the regime $E>\mu B$, we now rewrite equation (7) in the following form, which is the modified formula for Larmor precession calculated by the explicit time-evolved solution of the spatial parts, $\psi^{+}$ and $\psi^{-}$
\begin{equation}
\label{finalwfn}
\left|\Psi_{f}\right\rangle= \frac{A e^{i k x}}{\sqrt{2}} (c e^{i \phi_{1}}\left|\uparrow\right\rangle + b e^{i \phi_{2}}\left|\downarrow\right\rangle)\equiv \psi_{0}\chi(\phi)
\end{equation}
Here 
\begin{equation}
\label{c}
c=\sqrt{Re(\psi^{+}_{T})^{2}+Im(\psi^{+}_{T})^{2}}
\end{equation}
\begin{equation}
\label{d}
b=\sqrt{Re(\psi^{-}_{T})^{2}+Im(\psi^{-}_{T})^{2}}
\end{equation}
\begin{equation}
\label{e}
\phi_{1}= \tan^{-1}\frac{Im(\psi^{+}_{T})}{Re(\psi^{+}_{T})}
\end{equation}
\begin{equation}
\label{f}
\phi_{2}= \tan^{-1}\frac{Im(\psi^{-}_{T})}{Re(\psi^{-}_{T})}
\end{equation}

where, by using equations (\ref{psi-t}) and (13), we find that
\begin{eqnarray}
\label{repsi-t}
&&Re(\psi^{-}_{T})= \\
\nonumber
&&\frac{8k k_1(k^2+k_1^2)\sin(k a)\sin(k_1 a)+16k^2 k_1^2\cos(k a)\cos(k_1 a)}{(k+k_1)^4+(k-k_1)^4-2(k+k_1)^2(k-k_1)^2 \cos(2 k_1 a)}
\end{eqnarray}
\begin{eqnarray}
\label{impsi-t}
&&Im(\psi^{-}_{T})=\\
\nonumber
&&\frac{8k k_1(k^2+k_1^2)\cos(k a)\sin(k_1 a)-16k^2 k_1^2\sin(k a)\cos(k_1 a)}{(k+k_1)^4+(k-k_1)^4-2(k+k_1)^2(k-k_1)^2 \cos(2 k_1 a)}
\end{eqnarray}

\begin{eqnarray}
\label{repsi+t}
&&Re(\psi^{+}_{T})= \\
\nonumber
&&\frac{8k k_2(k^2+k_2^2)\sin(k a)\sin(k_2 a)+16k^2 k_2^2\cos(k a)\cos(k_2 a)}{(k+k_2)^4+(k-k_2)^4-2(k+k_2)^2(k-k_2)^2 \cos(2 k_2 a)}
\end{eqnarray}
\begin{eqnarray}
\label{impsi+t}
&&Im(\psi^{+}_{T})=\\
\nonumber
&&\frac{8k k_2(k^2+k_2^2)\cos(k a)\sin(k_2 a)-16k^2 k_2^2\sin(k a)\cos(k_2 a)}{(k+k_2)^4+(k-k_2)^4-2(k+k_2)^2(k-k_2)^2 \cos(2 k_2 a)}
\end{eqnarray}
\section{Limits of validity of the standard formula for Larmor precession}
Let us now examine in what limit the above expressions do indeed reduce to the standard expressions for Larmor precession. As already mentioned, we are working in the range $E>\mu B$. Let us now consider the stronger limit where $E\gg\mu B$. In this limit, the kinetic energy term of the Hamiltonian is appreciably larger than the potential energy term, and then, effectively, the time evolution of the entire wave function occurs due to a very shallow well and a very low barrier. This situation would correspond to the entire wave being transmitted, but picking up a phase. From the expressions for $k, k_1, k_2$, in the limit $E\gg\mu B$, we find that $k\approx k_1\approx k_2$.\\
 
In order to get the standard expression for Larmor precession, we will first set $k=k_1=k_2$ in Eqs.(\ref{repsi-t}),(\ref{impsi-t}), (\ref{repsi+t}), and (\ref{impsi+t}) except when they appear inside sine or cosine functions, since the latter terms are much more sensitive to the differences in values of $k, k_1, k_2$. Eqs. (\ref{repsi-t}) and (\ref{impsi-t}) then simplify to  
\begin{equation}
Re(\psi^{-}_{T})= \cos(k_1a-k a)
\end{equation}
\begin{equation}
Im(\psi^{-}_{T})=\sin(k_1 a - k a)
\end{equation}
Using the above expressions in Eqs. (\ref{d}) and (\ref{f}), we find that $b=1$ and $\phi_2=(k_1-k)a$. Similarly, rewriting equations (\ref{repsi+t}) and (\ref{impsi+t}), and using it in Eq.(\ref{c}) and (\ref{e}), we get $c=1$ and $\phi_1=(k_2-k)a$. Therefore Eq.(14) now has the form
\begin{equation}
\label{psiwfn1}
\left|\Psi_{f}\right\rangle= \frac{Ae^{i k x}}{\sqrt{2}} ( e^{i (k_2-k)a}\left|\uparrow\right\rangle + e^{i (k_1-k)a}\left|\downarrow\right\rangle)
\end{equation}
Since we have already stipulated the condition $k\approx k_1\approx k_2$, we can binomially expand $k_1$ and $k_2$ around $k$ and keep terms to the order of $\frac{\mu B}{E}$. Then $(k_1-k)a= - k\frac{\mu B}{2 E}a$. Using the relations $k=\frac{\sqrt{2 m E}}{\hbar}$ and $v=\frac{\hbar k}{m}$, we can write $(k_1-k)a=  \frac{\mu B}{\hbar}\frac{a}{v}=  \omega t$. Similarly, $(k_2-k)a= - \frac{\mu B}{\hbar}\frac{a}{v}=  -\omega t$. Therefore, we can write Eq.(\ref{psiwfn1}) as
\begin{eqnarray}
\left|\Psi_{f}\right\rangle&=& \frac{A e ^{i k x}}{\sqrt{2}} ( e^{- i \omega t}\left|\uparrow\right\rangle + e^{i \omega t}\left|\downarrow\right\rangle)\\
\nonumber
&&=\psi_{0}\frac{e^{- i\phi/2}}{\sqrt{2}}\left(\left|\uparrow\right\rangle_{z}+ e^{i\phi}\left|\downarrow\right\rangle_{z}\right)\equiv e^{- i\phi/2}\psi_{0} \chi(\phi)
\end{eqnarray}
which is exactly the equation we get from the spin-only treatment of the problem.

The above treatment brings out the curious feature that while the standard treatment ignores the kinetic energy term of the Hamiltonian, on solving the problem in a more complete manner, the same expression can be derived in the other extreme limit where the kinetic energy term is much higher than the spin-magnetic field interaction energy term.
\section{Quantitative estimates of the departure from the Larmor formula}
In the previous section we have shown that it is only when the kinetic energy associated with the wave function is much larger than the potential energy term in the Hamiltonian, in other words, the height of the potential barrier, or the depth of the potential well, we get back the standard expression for Larmor precession. However, when the kinetic energy term becomes comparable to the potential energy term due to the spin-magnetic field interaction, the standard expression no longer holds. In this section we calculate the effect of this deviation in terms of an observable, given by, in our case, the number of particles measured to be along $\left |\uparrow \right \rangle_{\theta}$ when the state emerging from the region of the magnetic field is passed through a Stern-Gerlach arrangement which is oriented at an angle $\theta$ with respect to the $+\hat x$ axis.

The state $\left |\uparrow \right \rangle_{\theta}$ is defined in the following manner $|\uparrow \rangle_{\theta}
=1/{\sqrt{2}}~\left(|\uparrow 
{\rangle}_z + e^{i \theta}|\downarrow {\rangle}_z \right)
$
The spin part of our original wave function is given by $\chi (0)=1/{\sqrt{2}}~ \left(|\uparrow {\rangle}_z +|\downarrow {\rangle}_z \right)$. According to the standard treatment of Larmor precession, the final spin state is given by $\chi (\phi)=1/{\sqrt{2}}~\left(|\uparrow {\rangle}_z + e^{i \phi}
|\downarrow {\rangle}_z \right)$. Therefore the probability of getting the particles with $\left |\uparrow \right \rangle_{\theta}$ is given by

\begin{equation}
\label{pplus}
p_{+}(\theta)={|}_{\theta}{\langle \uparrow}~|~\chi(\phi)\rangle|^2=\cos^2 (\theta-\phi)/2
\end{equation}
However, in the light of the complete treatment presented in the last section, the final spin part of the wave function is given by Eq. (14) to be $\chi^{\prime} (\phi)=1/{\sqrt{2}}~\left(c e^{i \phi_1}|\uparrow {\rangle}_z + b e^{i \phi_2}
|\downarrow {\rangle}_z \right)$. Consequently, the probability of getting particles with $\left |\uparrow \right \rangle_{\theta}$ is then modified which is of the form
\begin{equation}
\label{pplusprime}
p^{\prime}_{+}(\theta)=\frac{1}{2}\left(c^2+b^2+2 c b \cos(\phi_1-\phi_2+\theta)\right)
\end{equation}
It is clearly seen from Eqs.(\ref{pplus}) and (\ref{pplusprime}) that they are not same. We shall study the condition when they will be same. In the table given below, we compare the values of $p_{+}(\theta)$ and $p^{'}_{+}(\theta)$ for $\theta=0$ for different regimes of the velocity of the incident neutrons and the strength of the applied magnetic field. We can see clearly from the results given in the Tables 1 and 2 that when the incident velocity of the neutrons, or their kinetic energy is large, and the magnetic field is weak, $p_{+}(\theta)$ and $p^{\prime}_{+}(\theta)$ are the same. However, on increasing the strength of the magnetic field, or decreasing the velocity of the incident neutrons, there is an empirically verifiable difference between $p_{+}(\theta)$ and $p^{\prime}_{+}(\theta)$.

\begin{table}
\begin{tabular}{|c|c|c|}
\hline
$v(m/sec)$ & $p_{+}(\theta) $ & $p^{\prime}_{+}(\theta)$\\
\hline
\hline
$2000$ &0.40725&0.40725\\
\hline
$200$&0.645427&0.645464\\
\hline
$50$&0.690242& 0.653428\\
\hline
$10$&0.964184&0.855380\\
\hline
\end{tabular}
\caption{ \footnotesize Table 1: This table shows the numerical values of $p_{+}(\theta)$ and $p^{\prime}_{+}(\theta)$ for a fixed magnetic field, in this case, $2T$, while decreasing the velocity of the incident neutrons. For thermal neutrons, we see that the two are the same, while differences start arising for cold neutrons and this difference is appreciable for ultra-cold neutrons.}
\end{table}

\begin{table}
\begin{tabular}{|c|c|c|}
\hline
$B(Tesla)$ & $p_{+}(\theta) $ & $p^{\prime}_{+}(\theta)$\\
\hline
\hline
$0.5$ &0.997736&0.912933\\
\hline
$0.1$&0.645427&0.660003\\
\hline
$0.01$&0.407245& 0.407230\\
\hline
$0.001$&0.949661&0.949661\\
\hline
\end{tabular}
\caption{\footnotesize Table 2: This table shows the numerical values of $p_{+}(\theta)$ and $p^{\prime}_{+}(\theta)$ for a fixed incident velocity of neutrons, in this case, in the ultra-cold neutron regime of $10 m/s$, while decreasing the applied magnetic field.For a strong magnetic field, there is an appreciable difference between the two columns, which decreases as we decrease the magnetic field, till at low values of the field, the two are essentially the same.}
\end{table}
\section{Generalization of the Larmor precession treatment for calculating the spin distribution for a wave packet}
We will now use the results derived above to analyze the spin distribution which arises due to spin-magnetic field interaction when a wave packet passes through the spin rotator arrangement. For this purpose, we use a Gaussian wave packet as the space part of our incident wave function, instead of a plane wave as was used in Sections II and III. We choose the spin-polarization of the incident wave function to be in the $+x$ direction, and the magnetic field to be pointing along the $+z$ direction. Thus our incident wave function is given by
\begin{eqnarray}
\left|\Psi_i\right\rangle=\frac{1}{(2 \pi \delta ^2)^{\frac{1}{4}}}e^{-\frac{(x-x_0)^2}{4 \delta ^2}}e^{i k_0 x}\frac{1}{\sqrt{2}}\left(\left|\uparrow\right\rangle+\left|\downarrow\right\rangle\right)
\end{eqnarray}
where $x_0$ is the initial peak of the wave packet, $k_0$ is the peak wave-number and $\delta$ is the width of the incident wave packet. In the previous section we have seen that the precession of spin caused by interaction with the magnetic field within a spin rotator of given parameters is a function only of $k$. Therefore, while dealing with the Gaussian wave packet, it is convenient to use the  Fourier transform of the incident wave function, given by
\begin{equation}
\left|\Psi_i\right\rangle=\left(\frac{2 \delta^2}{\pi}\right)^{\frac{1}{4}}e^{-\delta ^2(k-k_0)^2}e^{i k x_0}\frac{1}{\sqrt 2}\left(\left|\uparrow\right\rangle+\left|\downarrow\right\rangle\right)
\end{equation}
Using results from the previous section, we can then write the final wave function in the Fourier basis to be
\begin{eqnarray}
\left|\Psi_f\right\rangle&=&\left(\frac{2 \delta^2}{\pi}\right)^{\frac{1}{4}}e^{-\delta ^2(k-k_0)^2}e^{i k x_0}\\
\nonumber
&&\times\frac{1}{\sqrt{2}}\left(c(k)e^{i\phi_1(k)}\left|\uparrow\right\rangle+b(k)e^{i\phi_2(k)}\left|\downarrow\right\rangle\right)
\end{eqnarray}
where $c, b, \phi_1, \phi_2$ are as defined in Eqs. (15),(16),(17), and (18), and hence have different values for different values of $k$. From the above equation, it becomes clear that we have a spin distribution which occurs as a result of the spin-magnetic field interaction of different wave-number components of the original wave packet. 

We will now find the distribution of spins along $\left|\chi\right\rangle=\frac{1}{\sqrt{2}}\left(\left|\uparrow\right\rangle+\left|\downarrow\right\rangle\right)$ or the $+x$ direction. The projection of the spin of the final wave function along this direction is given by
\begin{eqnarray}
\langle\chi\left|\Psi_f\right\rangle&=&\left(\frac{2 \delta^2}{\pi}\right)^{\frac{1}{4}}e^{-\delta ^2(k-k_0)^2}e^{i k x_0}\\
\nonumber
&&\times\frac{1}{2}\left(c(k)e^{i\phi_1(k)}+b(k)e^{i\phi_2(k)}\right)
\end{eqnarray}
Therefore, the probability of finding spins along the $+x$ direction will be given by
\begin{eqnarray}
\left|\langle\chi\left|\Psi_f\right\rangle\right|^2=&&\left(\frac{2 \delta^2}{\pi}\right)^{\frac{1}{2}}e^{-2 \delta ^2(k-k_0)^2}\\
\nonumber
&&\times\frac{1}{4}\left(c(k)e^{i\phi_1(k)}+b(k)e^{i\phi_2(k)}\right)\\
\nonumber
&&\times\left(c^{*}(k)e^{-i\phi_1(k)}+b^{*}(k)e^{-i\phi_2(k)}\right)
\end{eqnarray} 
Such a spin probability function can be measured by a Stern Gerlach arrangement for the particles emerging from the spin rotator. A crucial feature to be stressed here is that the above spin probability function given by Eq. (33) has been obtained from our complete treatment of the problem of time evolution of the spin of a particle passing through a region of uniform magnetic field. On the other hand, the counterpart of such a spin probability distribution can also be obtained from the standard treatment of Larmor precession. A comparison between the results obtained from these two different approaches for various values of the magnetic field and velocity of the incident neutrons is illustrated in Figures 1 and 2. Similar to the case of the plane wave, we notice that the distributions from the two different approaches overlap when the incident velocity of the neutrons is high and the magnetic field is weak, whereas a noticeable divergence appears upon either decreasing the incident velocity or increasing the strength of the magnetic field.
\begin{figure}[h]
{\rotatebox{0}{\resizebox{7.0cm}{3.0cm}{\includegraphics{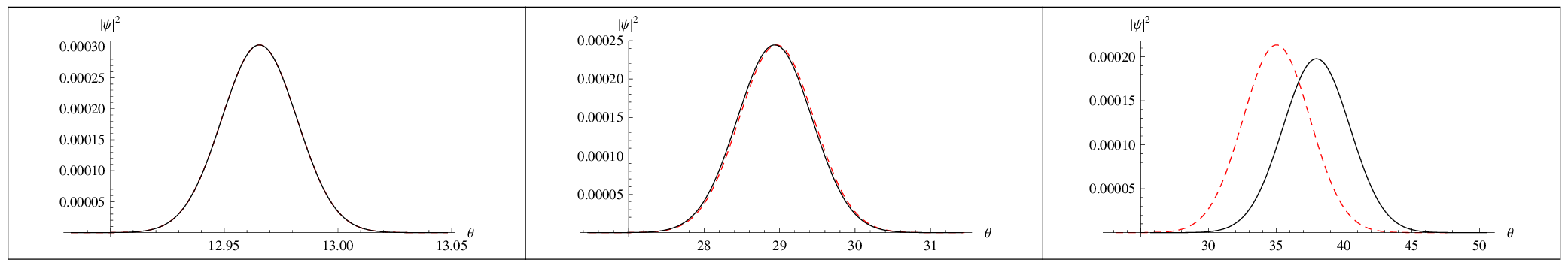}}}}
\caption{\footnotesize (Color online) Here in the successive plots, we show the resultant probability distribution of spins emerging from the spin rotator, calculated according the standard Larmor precession formula (lighter red curve) and the modified formula (darker black curve) given in this paper, while varying the constant magnetic field applied in the region of the spin rotator. The incident velocity is fixed at $10m/s$, while the applied magnetic field takes the values $0.001T$, $0.03T$ and $0.15T$. The observable effect of the departure from the standard expression becomes more pronounced as the magnetic field is increased.}
\end{figure}
\begin{figure}[h]
{\rotatebox{0}{\resizebox{7.0cm}{3.0cm}{\includegraphics{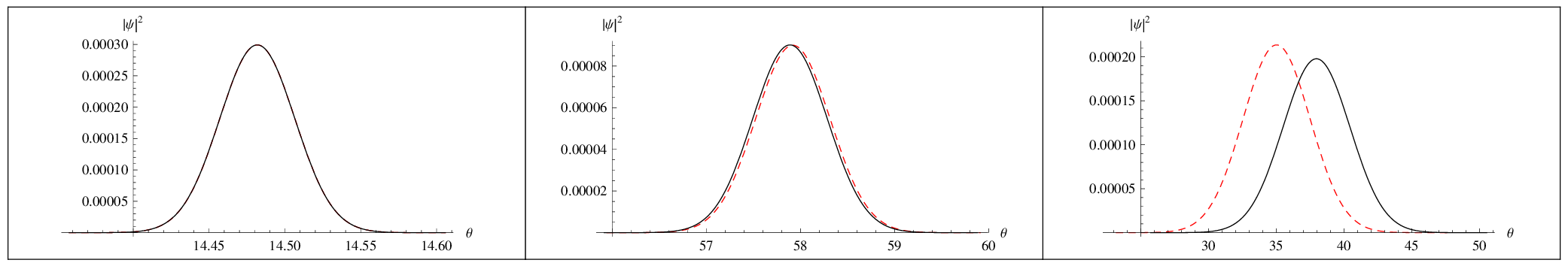}}}}
\caption{\footnotesize (Color online) Here in the successive plots, we show the resultant probability distribution of spins emerging from the spin rotator, calculated according the standard Larmor precession formula (lighter red curve) and the modified formula (darker black curve) given in this paper, while varying the incident velocity of the neutrons. The value of the applied magnetic field is held at $0.15T$, while the incident velocities are taken to be $100m/s$, $25m/s$ and $10m/s$. The observable effect of the departure from the standard expression becomes more pronounced when the incident velocity of the neutrons is decreased.}
\end{figure}

\section{conclusion and outlook}
In a nutshell, the central result of the present paper is that the standard expression for Larmor precession holds true only in the regime where the kinetic energy of the system is much greater than the potential energy term arising out of the interaction between the spin of the particle and the magnetic field. It is essentially when these two terms become comparable that the departure from the standard expression for Larmor precession becomes appreciable to the extent of being empirically observable - such a departure can indeed be tested by choosing appropriate conditions of high magnetic field and low velocity of incident neutrons; for example, in the experiments using cold neutrons.  

An interesting application of the above treatment could be in enabling the construction of an effective transit time distribution for a spin-polarized wave packet passing through a spin-rotator containing uniform magnetic field, of course, subject to the constraint of choosing the relevant parameters appropriately such that the rotation of spin pertaining to any wave-number component of the wave packet does not exceed $\pi$. A detail derivation of the spin probability distribution emerging from a spin-rotator by following our treatment that enables identifying precisely the regime in which the standard formula of Larmor precession is valid would thus be a crucial ingredient for using a spin-rotator as a quantum clock \cite{pan}. Such a clock may particularly be useful for measuring the arrival/transit time distributions, and for making a quantitative study of the possible differences in the predictions obtained from the different quantum mechanical schemes suggested for computing the arrival/transit time distributions \cite{mugarev,mugabook}. Further studies along this line using the exact formula for Larmor precession derived in this paper are called for, the results of which could be compared with other models for quantum clock suggested in the literature \cite{wigner}. 

Among other possible uses of the exact formula for Larmor precession in the light of the recent significant experiments, here we may mention, for example, the neutron interferometric experiment \cite{yuji} testing single particle Bell's inequality \cite{basu} involving entanglement between the path and the spin degrees of freedom of a spin-1/2 particle. In such an experiment, the spin flipper that is placed in one of the two paths of the interferometer plays a crucial role in generating the path-spin entangled state, since the spin flipper is ideally required to flip spins of all the neutrons passing through it, so that the flipped state and the unflipped spin state in the respective two paths are completely orthogonal. In order to ensure this condition, the choice of the relevant parameters has to be carefully made on the basis of an appropriate formula for Larmor precession. Usually this is done by using the standard Larmor formula, as in the above mentioned experiment \cite{yuji}. It is in such context that the exact formula for Larmor precession obtained in this paper could also be useful.

{\it Acknowledgements:} We are grateful to H. Rauch, A. Sudbery, V. Scarani, Y. Hasegawa, A. Matzkin and T. S. Mahesh for useful discussions based on an earlier version of this work. DH acknowledges support from the DST Project SR/S2/PU-16/2007. DH also thanks the Centre for Science, Kolkata for support. AKP acknowledges the support from JSPS Postdoc Fellowship for Foreign Researcher and Grant-in-Aid for JSPS fellows No. 24-
02320.

\end{document}